\documentclass[prl,twocolumn,superscriptaddress]{revtex4}
\usepackage{amsmath,amssymb}
\usepackage{graphics,graphicx}
\usepackage{dcolumn,bm}
\usepackage{psfrag}
\usepackage[utf8x]{inputenc}
\newcommand{\icf}
{\affiliation{Instituto de Ciencias F\'isicas, Universidad Nacional Aut\'onoma de M\'exico, Cuernavaca, M\'exico }}

\newcommand{\cic}
{\affiliation{Centro Internacional de Ciencias, Cuernavaca, Mexico}}

\begin{document}

\title{Correlation Matrix Spectra: A Tool for Detecting Non-apparent Correlations?}

\author{Soham Biswas}%
 \email{soham@fis.unam.mx}
\icf 

 \author{Francois Leyvraz}%
 \email{leyvraz@fis.unam.mx}  
 \icf
 \cic

\author{Paulino Monroy Castillero}%
  \email{espaulino@gmail.com }
\icf 

\author{Thomas H Seligman}%
 \email{seligman@icf.unam.mx }
\icf
\cic

\begin{abstract}

It has been shown that, if a model displays long-range (power-law) spatial correlations, its equal-time correlation matrix 
of this model will also have a power law tail in the distribution of its high-lying eigenvalues. The purpose of this letter 
is to show that the converse is generally incorrect : a power-law tail in the high-lying eigenvalues of the correlation 
matrix may exist even in the absence of equal-time power law correlations in the original model. We may therefore view
the study of the eigenvalue distribution of the correlation matrix as a more powerful tool than the study of correlations,
one which may in fact uncover structure, that would otherwise not be apparent. Specifically, we show that in the Totally 
Asymmetric Simple Exclusion Process, whereas there are no clearly visible correlations in the steady state, the eigenvalues of
its correlation matrix exhibit a rich structure which we describe in detail. 
\end{abstract}

\maketitle

The analysis of correlation matrices has attracted considerable attention almost for a 
hundred years starting with multivariate analysis in finance \cite{finance}. 
In two pioneering papers Laloux {\em et al.} and Plerou {\em et al.}
analysed a complex time signal---a time series of stock prices---and successfully disentangled 
the part due to chance and the systematic part via an analysis of the eigenvalues of
the correlation  matrix \cite{laloux,plerou,stanley}. 
The same tools of correlation matrix analysis have recently
gained attention from physicists in the discussion of critical 
phenomena and phase transitions: If we consider an extended system undergoing
some kind of dynamics, the equal time correlations between the various components of the system
yield a correlation matrix, the eigenvalues of which can be analysed.
In this context, it has recently been shown \cite{prosen}, that a 
power-law decay of correlations in space leads to a power-law behaviour for  the large eigenvalues 
of the correlation matrix . 
We are thus led to ask whether the opposite is true. 
That is, does the observation of such power-law behaviour in the eigenvalues imply a power-law in 
spatial correlations?

In a trivial sense, it is possible to find systems for which no correlations are apparent,
and yet the power-law behaviour of the eigenvalues remains: We simply take a system which does display 
spatial power-law correlations, and ``scramble'' the components by randomly permuting them. 
Such an operation leaves the eigenvalues invariant, so that their
power-law behaviour testifies to the existence of the spatial correlations, even though the latter have
been masked by the random permutation.

However, we may ask whether there exist less trivial counterexamples.
In the following, we shall suggest that there probably are: 
we shall analyse the so-called Totally Asymmetric Simple Exclusion Process,
(TASEP) which shows little or no apparent spatial correlation, which additionally surely does not
have a power-law decay, and yet, as we report here, 
displays a marked power-law feature
in the spectra of its correlation matrix on one critical line of its phase diagram as well as further
anomalous structure in the rest of the phase diagram. Of course, it
is difficult rigorously to exclude the possibility that non-trivial spatial correlations for this system
have, in fact, been hidden by a ``scrambling'' process similar to that described in the 
last paragraph, but, in view of the system's simplicity, this does not seem likely.

TASEP is a model consisting of a many-particle hopping system where particles are located on a discrete lattice that evolves
in continuous time. Particles can hop to the next lattice site, in only one direction (say to the right-hand side), on a 
one-dimensional lattice at a random time, with rate one, provided that the target site is empty. We here consider the problem 
with open boundary conditions where both sides of the lattice are coupled with particle reservoirs. If the first site of the 
lattice is empty then a particle can hop from the reservoir into the system with a transition rate $\alpha$ and the 
particles leave the system from the last site of the lattice with a transition rate $\beta$. TASEP has been used to
describe directed transport in 1D, such as arises, for instance, in unidirectionally moving vehicular traffic 
along roads \cite{road, rev}. 

\begin{figure}[ht]
\includegraphics[width=6.5cm,angle=0]{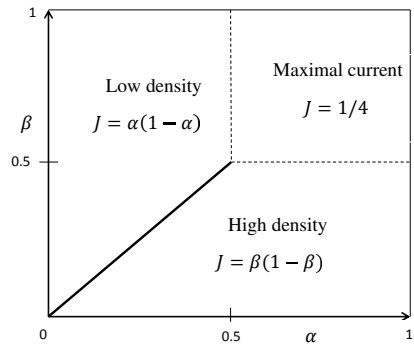}
\caption{Phase diagram of TASEP with open boundary conditions at the thermodynamic limit, consisting
of the high-density phase (I), the  low-density phase (II) and the maximum-current phase (III). }
\label{sch_tasep}
\end{figure}

There are several reasons to choose TASEP for the present study:  
The equal-time correlation functions for this 
stationary non-equilibrium system are  known exactly \cite{tasep_exact, Derrida} and the phase 
diagram (see Fig.~\ref{sch_tasep}) of the exact and the mean field solutions coincide. In phase I, there
is a high density of occupied sites which fluctuates little in time, in phase II there is a corresponding low-density phase,
in which the density is also approximately time-independent, whereas in III the density is equal to $1/2$, 
independent of $\alpha$ and $\beta$. Finally, in the transition line between I and II, a phase exists in which the density
oscillates between a high value corresponding to a nearby point of I, and a low value corresponding to  a
nearby point of II. 

\begin{figure}[ht]
\includegraphics[width=8.4cm,angle=0]{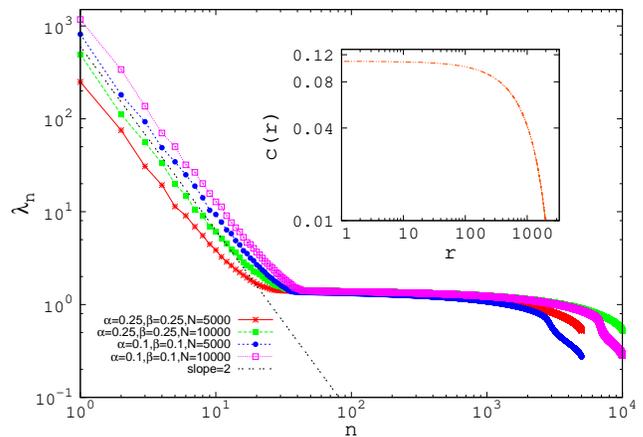}
\caption{The Zipf  plot for the ranked eigenvalues for different values of $\alpha$ and $\beta$ on the low density-high 
density coexistence line ($\alpha=\beta$ line). Inset shows the decay of spatial two point correlation function $C(r)$ 
with distance $r$. }
\label{sp_power}
\end{figure}

We are interested in the density-density two point correlation $C_{i,j} $ on the lattice. 
This is defined as the probability to find a particle in a lattice site 
$j$, given there is a particle at lattice site $i$. 
The analytical expression for this correlation function $C(r)$ (average correlation 
between any two points at distance $r$) is given in  
\cite{Derrida}. On the line $\alpha=\beta < 0.5$, where both the high density and low 
density phase coexist, the two point correlation function do not decay 
in a power law fashion. Rather, the spatial correlations appear to decay
at a scale of the order of the system size, as shown in the inset of Figure \ref{sp_power}.

In the present letter we analyse the spectrum of the correlation 
matrix and particularly, its behaviour as $\alpha$ and $\beta$ vary.
For  $\alpha=\beta < 0.5$ we find a power-law (Fig. \ref{sp_power}) in the Zipf plot (see below)
and thus an example, where we find a 
power-law even though the obvious two-point function in space 
does not show such a behaviour. 
Let us now compare the eigenvalue density of the correlation matrix 
with the null-hypothesis, that is, the eigenvalue spectrum of the correlation matrix
of a completely uncorrelated signal. 
These correlations remain different from zero if 
they are taken on the scale of the fluctuations, that is, on the scale of the square root of 
the duration of the signal. The eigenvalue distribution may be calculated exactly in this case, and the eigenvalue 
density was determined analytically by Mar\v cenko and Pastur \cite{MP}. This eigenvalue density
has the remarkable feature that it vanishes outside a finite interval. We may thus meaningfully speak of
deviations from the Mar\v cenko--Pastur (MP) result whenever eigenvalues appear significantly outside 
this interval. 

We thus compare our eigenvalue spectra with the MP  distribution in order to see 
to what extent our eigenvalues differ from an uncorrelated signal. This is very much 
in the line of \cite{laloux,plerou}. Our key result can now be stated as follows:
we find agreement in the high and low density regions, that is, in the interior of 
regions I and II, so that these are indeed well described by a random process. 
On the other hand, in all other parts of the phase diagram, characteristic differences are observed: 
in the constant current phase corresponding to III, we find a 
significant deviation from the MP  prediction in that a significant number of
eigenvalues below the MP threshold are observed. We find similar
deviations on the I/III and II/III lines, and different ones at the triple point $\alpha=\beta=1/2$.
 
To construct the correlation matrices, we have generated the 
times series by Monte-Carlo simulation. The random update rule was been
used to generate the time series. The lattice size is $N$, with $10^3 \leq N \leq 10^4$. 
For each parameter value we have considered the length of the 
time series as $T=20 \times N$. Obviously the correlation matrix $C$ is an $N \times N$ 
dimensional matrix. The results are averaged over an ensemble of $100$ configurations. 
 
We shall analyse the eigenvalues using the so called Zipf  plot, also known 
as ``scree diagram'' or ranking-of-eigenvalues plot, 
in which the eigenvalues in decreasing order
$\lambda_n$ are plotted against their rank $n$, typically on 
a doubly logarithmic plot. Such a plot makes an initial power-law 
very prominent. 

\begin{figure}[ht]
\includegraphics[width=6cm,angle=270]{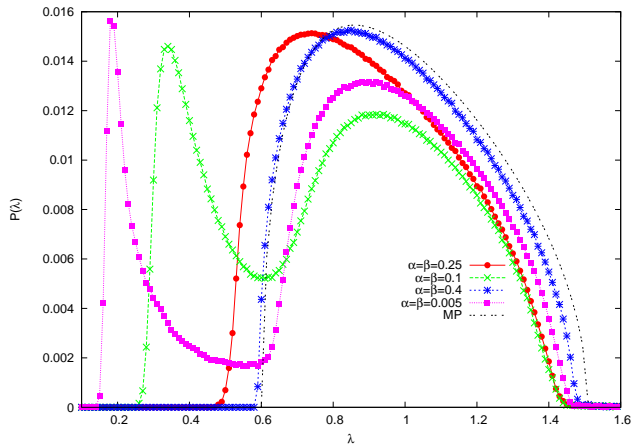}
\caption{The plot of bulk for the distribution of eigenvalues for different values of $\alpha$ and $\beta$ on the 
low density-high density coexistence line ($\alpha=\beta < 0.5$ line). The MP  distribution is shown by the 
black double dashed line.}
\label{sp}
\end{figure}

Let us first look at the structure of eigenvalues on the I/II coexistence line, that is, for  $\alpha=\beta < 0.5$. For 
any value of $\alpha$ and $\beta$ on this low density-high density
coexistence line, though the spatial two point correlation function $C(r)$ decay with distance $r$, 
but do not decay in a power law [see inset of the  figure \ref{sp_power}]. So the density 
of eigenvalues of the correlation matrix will obviously be 
different from the MP  distribution  [Fig. \ref{sp}]. We do find an initial 
power-law decay on the Zipf plot for the eigenvalues on this coexistence line.
The power we find ($\lambda_n \sim n^{-\theta}$ with $\theta \approx 2$) obtained  is same for any value of 
$\alpha$ and $\beta$, as long as $\alpha=\beta < 0.5$ [Fig. \ref{sp_power}]. 

There are other differences between the  observed distribution on the I/II separation line and the
MP  distribution: first, the range over which the power-law is observed, varies 
with the parameter value. It is higher for the lower 
values of $\alpha=\beta$. As a result, the density of 
high-lying eigenvalues
differ for different values of $\alpha$ and $\beta$ on this line. Second, we observe a 
shift of the bulk to lower eigenvalues in compared with the MP  
distribution [Fig.~\ref{sp}]. This shift actually compensates the contribution from the higher eigenvalues,  since 
the sum of the eigenvalues remains constant and equal to the dimension of the matrix. However for lower values of $\alpha$, 
the density profile for the eigenvalues are deformed and the deformation becomes more prominent as the value of 
$\alpha= \beta$ decreases [Fig. \ref{sp}]. Finally, when $\alpha\lesssim 0.25$, a second peak appears in the eigenvalue
density, in sharp contrast to the MP result, in which only one peak appears.

This phenomenon may perhaps be explained as follows: the density of particles inside the lattice, in the low 
density region is lower for the 
lower values of $\alpha$. Similarly it is higher for the lower values of $\beta$, in the high 
density region. Hence for $\alpha=\beta < 0.5$, that is on the 
I/II coexistence line, for the lower values of $\alpha$ and $\beta$ there will be larger strings of particles 
in the lattice followed by a string of empty lattice sites of similar length. As a result correlation 
length inside the lattice increases as $\alpha$ decreases (for $\alpha=\beta < 0.5$). This correlation length is not enough 
to show a power law decay in case of two point correlation 
function, but it may well be related to the presence of a larger number eigenvalues above the MP threshold. These
then display the power law behaviour observed in the Zipf plot. 

We now proceed to consider the new peak in the eigenvalue density that appears for $\alpha=\beta\lesssim0.25$.
In particular, it is 
natural to ask whether, inside the two peaks of eigenvalue density which arise for 
lower values of $\alpha$ and $\beta$, the correlations of the unfolded eigenvalues (say $\xi$) \cite{mehta}, are identical. 
We test two independent statistical properties of unfolded eigenvalues $\xi$: the distribution of nearest-neighbour
spacing $s= \xi_{i+1} - \xi_i$ and the statistics of number variance $\Sigma^2(x)$. 
The distributions of nearest-neighbour spacing of unfolded eigenvalues, which are obtained from the first and the second peak of the density 
of eigenvalues [Fig. \ref{sp}] appear to be universal. That means the behaviour of the nearest-neighbour (nn) spacing distributions are not 
distinguishable from that of the Wishart ensemble for both the peaks. 

\begin{figure}[ht]
\includegraphics[width=8.5cm,angle=0]{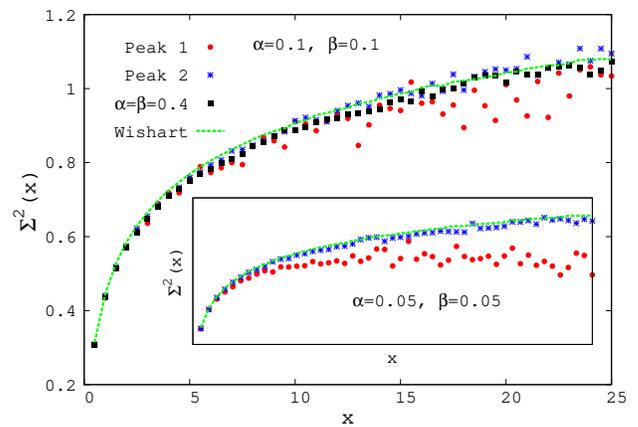}
\caption{Number variance $\Sigma^2 (x)$, is plotted against the interval length $x$, calculated separately from peak 1 (red points) and 
peak 2 (blue stars) of their eigenvalue density for points $\alpha = \beta=0.1$ and 
$\alpha = \beta=0.05$. For $\alpha = \beta=0.4$ (black squares)
 $\Sigma^2 (x)$ is calculated from the single bulk of its eigenvalue density. Continuous line is 
 of a  Wishart ensemble of $2000$ configurations, plotted for comparison.}
\label{numb_var}
\end{figure}

{\sl Number variance}, the variance of the number of unfolded eigenvalues in the intervals of length $x$, is defined as
$\Sigma^2(x) =\langle [n_\xi(x)-x]^2 \rangle_\xi$,
where  $n_\xi(x)$ is the number of unfolded eigenvalues in the interval $[\xi-x/2,\xi+x/2]$. The average is made along $\xi$.
If the eigenvalues are uncorrelated, $\Sigma^2(x) = x$; whereas if all unfolded eigenvalues are 
equidistant, $\Sigma^2(x) =0$.
We found the unfolded eigenvalues obtained from the first peak of the distribution [Fig \ref{sp}] does not follow the universal behavior 
for the number variance statistics, while for the second peak it appears to be universal [Fig. \ref{numb_var}].

\begin{figure}[ht]
\includegraphics[width=8.6cm,angle=0]{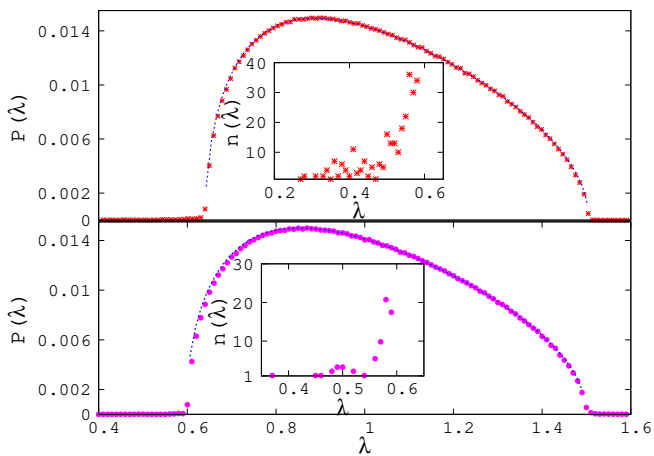}
\caption{The probability distribution of eigenvalues at the maximal current 
(upper panel) regime and for the triple point (lower panel). The MP  distribution 
is shown by the blue dotted line. Inset shows the number distribution of below threshold eigenvalues.}
\label{mc}
\end{figure}

In the maximal or constant current region, we also find significant deviations  from MP  
in  the probability distribution of eigenvalues [see Fig. \ref{mc}]. 
This consists in the appearance of eigenvalues below the lower threshold of MP, and is quite
pronounced.
One or two high eigenvalues (outside the limits of MP) are also observed. 
The deviation of the probability distribution of eigenvalues of the correlation matrix is also present 
on the transition lines of I/III and II/III. But there the number of below-threshold
eigenvalues distribution is smaller than in the constant 
current phase. 
The plot for the eigenvalue density for the maximal current and for the triple point is shown in Fig \ref{mc}. 
Transition from the MC phase to the triple point is continuous, as the number of lower 
eigenvalues decreases slowly as the triple point is approached.

\begin{figure}[ht]
\includegraphics[width=8.5cm,angle=0]{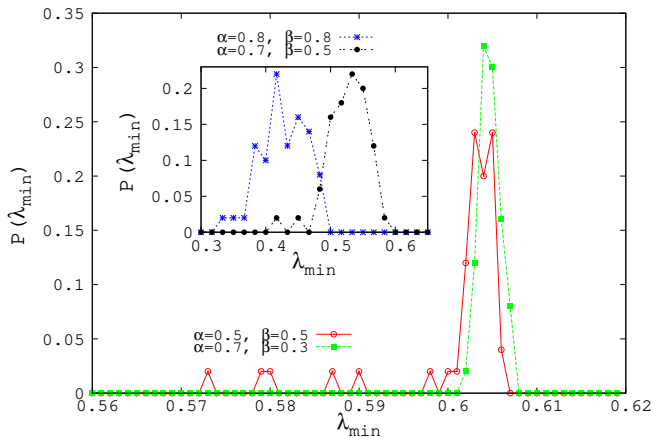}
\caption{(Color Online) Distribution of the lowest eigenvalues, over the 
configuration space, for different part of the phase diagram.
We have averaged over $100$ configurations. }
\label{loweigen}
\end{figure}

In different parts of the phase diagram, the probability densities of the eigenvalues have a deviation 
from the Mar\v cenko--Pastur but also the distribution of the lowest eigenvalues, over the 
configuration space is significantly different from that of the low density or 
high density regions [Fig. \ref{loweigen}]. 

We have also checked whether the effects can be accounted for by edge effects: we did 
not notice any significant such effect for any value of $\alpha$ and 
$\beta$ for the entire phase diagram. This is also true for the higher eigenvalues 
when $\alpha=\beta < 0.5$. If there is any edge effect at all 
in the spectrum of eigenvalues, it is not detectable with the present computational accuracy.

On the  I/II phase coexistence line, 
we have taken different parts of the lattice and repeated the correlation matrix analysis. We indeed observed 
the power law in the Zipf plot (with the same value of $\theta$) for all the parts of the lattice
which will be discussed in detail at \cite{unpub}. On this coexistence line the motion of the domain wall
is non-localised over the lattice \cite{domain}. Whether or not the motion of the domain wall is responsible for 
the observed power law will be studied as a future problem \cite{unpub}. 
We will also attempt to connect the formula of two point function given in \cite{Derrida} to the exact solutions derived recently \cite{gur1}
for arbitrary correlations at least in an approximate fashion.

In conclusion, we have shown that the analysis of the density of eigenvalues of the correlation matrix 
of a signal is sensitive to non-trivial correlations, which cannot otherwise be reliably characterised
by direct numerical observation. Such is the case of TASEP, in which two-body correlations are 
weak, though they extend over the whole system at the phase coexistence line. Comparison of the correlation matrix spectrum
with those generated by a random signal provide clear evidence 
that the signal produced by TASEP has significant correlation in some parts of the phase diagram. 
 In particular at the $\alpha = \beta <1/2$ line long-range weak correlations in space induce a power law in the spectrum.

Acknowledgement : The authors acknowledge
financial support from CONACyT through Project  Fronteras 102 and Program 
UNAM---DGAPA PAPIIT IN114014 as well as the use of Miztli supercomputer 
of DGTIC, UNAM for the computational resources under the 
project number SC15-1-S-20. S.B also acknowledges a postdoctoral fellowship from DGAPA/UNAM.

\end{document}